\documentclass[12pt,oneside,UTF8,AutoFakeBold=4]{article}

\usepackage[top=1in,textheight=8.8in,textwidth=7.1in]{geometry} 
\usepackage{bm,amssymb,amsfonts,amsthm,amscd,mathrsfs,mathtools}
\makeatletter
\renewcommand*\env@matrix[1][\arraystretch]{%
  \edef\arraystretch{#1}%
  \hskip -\arraycolsep
  \let\@ifnextchar\new@ifnextchar
  \array{*\c@MaxMatrixCols c}}
\makeatother
\usepackage{nicematrix}
\usepackage{empheq}
\usepackage{braket} 
\usepackage{enumitem}
\usepackage{outlines} 
\usepackage[font=itshape,begintext=``,endtext='']{quoting} 
\usepackage{tabu} 
\usepackage{booktabs} 
\usepackage{graphicx} 
\usepackage{subfigure}
\usepackage{grffile} 
\usepackage[margin=1.1cm, font=footnotesize]{caption} 
\usepackage{float} 
\usepackage{tikz}  
\usetikzlibrary{cd} 
\definecolor{NSFC_blue}{RGB}{0,112,192}
\definecolor{G_red}{RGB}{234,67,53}
\definecolor{G_yellow}{RGB}{251,188,5}
\definecolor{G_green}{RGB}{52,168,83}
\definecolor{G_blue}{RGB}{66,133,244}
\definecolor{G_blue2}{RGB}{0,72,204}
\definecolor{M_red}{RGB}{243,83,37}
\definecolor{M_yellow}{RGB}{255,186,8}
\definecolor{M_green}{RGB}{129,188,6}
\definecolor{M_blue}{RGB}{5,166,240}

\usepackage[backend=biber,style=authoryear,giveninits=true,uniquename=false,date=year,sortcites,natbib=true,sorting=none]{biblatex}
\usepackage{algorithm}
\usepackage[noend]{algpseudocode}

\usepackage{nomencl} 
\makenomenclature

\usepackage[unicode,hidelinks,colorlinks=true,citecolor = M_blue]{hyperref} 
\usepackage[nameinlink,capitalise]{cleveref} 
\crefname{enumi}{case}{cases}
\usepackage{titling} 
\thanksmarkseries{fnsymbol}
\usepackage{authblk} 
\newcommand{\rc}{\mathrm{c}}

\newcommand{\ri}{\mathrm{i}}

\newcommand{\rr}{\mathrm{r}}

\newcommand{\bA}{\mathbf{A}}

\newcommand{\bI}{\mathbf{I}}

\newcommand{\bL}{\mathbf{L}}
\newcommand{\bM}{\mathbf{M}}
\newcommand{\bN}{\mathbf{N}}

\newcommand{\bP}{\mathbf{P}}
\newcommand{\bQ}{\mathbf{Q}}
\newcommand{\bR}{\mathbf{R}}
\newcommand{\bS}{\mathbf{S}}

\newcommand{\bU}{\mathbf{U}}
\newcommand{\bV}{\mathbf{V}}
\newcommand{\bW}{\mathbf{W}}

\newcommand{\bma}{\bm{a}}
\newcommand{\bmb}{\bm{b}}
\newcommand{\bmc}{\bm{c}}
\newcommand{\bmd}{\bm{d}}
\newcommand{\bme}{\bm{e}}

\newcommand{\bmu}{\bm{u}}
\newcommand{\bmv}{\bm{v}}

\newcommand{\bbC}{\mathbb{C}}

\newcommand{\bbR}{\mathbb{R}}

\newcommand{\hatx}{\widehat{x}}

\newcommand{\hatz}{\widehat{z}}
\newcommand{\hatA}{\widehat{A}}

\newcommand{\hatbA}{\widehat{\mathbf{A}}}

\newcommand{\hatbN}{\widehat{\mathbf{N}}}

\newcommand{\tilA}{\widetilde{A}}

\newcommand{\tilbA}{\widetilde{\mathbf{A}}}

\newcommand{\bDelta}{\bm{\Delta}}

\newcommand{\bLambda}{\bm{\Lambda}}

\newcommand{\hatbGamma}{\widehat{\bm{\Gamma}}}

\theoremstyle{plain}
\newtheorem{theorem}{Theorem}[section]

\newtheorem{corollary}{Corollary}[section]

\theoremstyle{definition}

\theoremstyle{remark}
\newtheorem{remark}{Remark}[section]

\newcommand{\zero}{\mathbf{0}}
\newcommand{\ii}{\mathrm{i}}

\newcommand{\T}{\mathsf{T}} 
\newcommand{\diag}{\mathrm{diag}\,}
\newcommand{\tr}{\mathrm{tr}\,} 

\newcommand{\parentheses}[1]{\left(#1\right)}

\newcommand{\norm}[1]{\left\|#1\right\|}

\makeatletter
\newsavebox{\@brx}
\newcommand{\llangle}[1][]{\savebox{\@brx}{\(\m@th{#1\langle}\)}%
  \mathopen{\copy\@brx\kern-0.5\wd\@brx\usebox{\@brx}}}
\newcommand{\rrangle}[1][]{\savebox{\@brx}{\(\m@th{#1\rangle}\)}%
  \mathclose{\copy\@brx\kern-0.5\wd\@brx\usebox{\@brx}}}
\makeatother

\usepackage{stackengine}
\stackMath
\newcommand\tsup[2][2]{%
 \def\useanchorwidth{T}%
  \ifnum#1>1%
    \stackon[-1.3ex]{\tsup[\numexpr#1-1\relax]{#2}}{\mathaccent"0365{}\kern-.5pt}%
  \else%
    \stackon[-1ex]{#2}{\mathaccent"0365{}\kern-.5pt}%
  \fi%
}

\title{Schur Forms and Normal-Nilpotent Decompositions
\thanks{Will appear in \emph{Applied Mathematics and Mechanics} Vol. 45, No. 9, Sep. , 2024. This is an author-edited version.}}

\author[1]{Zhen LI 
\thanks{Correspondence email: lizhen0102@bimsa.cn}
}
\affil[1]{Beijing Institute of Mathematical Sciences and Applications}
\date{May 08, 2024}

\begin{document}
\maketitle
\graphicspath{{pics/}}

\abstract
Real and complex Schur forms have been receiving increasing attention from the fluid mechanics community recently, especially related to vortices and turbulence.
Several decompositions of the velocity gradient tensor, such as the triple decomposition of motion (TDM) and normal-nilpotent decomposition (NND), have been proposed to analyze the local motions of fluid elements.
However, due to the existence of different types and non-uniqueness of Schur forms, as well as various possible definitions of NNDs, confusion has spread widely and is harming the research.
This work aims to clean up this confusion.
To this end, the complex and real Schur forms are derived constructively from the very basics, with special consideration for their non-uniqueness.
Conditions of uniqueness are proposed.
After a general discussion of normality and nilpotency, a complex NND and several real NNDs as well as normal-nonnormal decompositions are constructed, with a brief comparison of complex and real decompositions.
Based on that, several confusing points are clarified, such as the distinction between NND and TDM, and the intrinsic gap between complex and real NNDs.
Besides, The author proposes to extend the real block Schur form and its corresponding NNDs for the complex eigenvalue case to the real eigenvalue case. 
But their justification is left to further investigations.
\section{Introduction}

In the thesis of \citet{liTheoreticalStudyDefinition2010a}, a normal-nilpotent decomposition (NND) of real square matrices is proposed and applied to the velocity gradient tensor of a fluid element for analyzing vortex criteria (also see \cite{liEvaluationVortexCriteria2014a,liEvaluationVortexCriteria2014b}).
The decomposition can be stated as that every 3-by-3 real matrix is a sum of a normal matrix and a nilpotent matrix.
As a special application, the velocity gradient tensor $\bA=\nabla\bmv$ can be decomposed as
\begin{align}
    \bA = \bN + \bS,
\end{align}
where $\bN$ is a normal tensor and $\bS$ is a nilpotent tensor.
Then $\bN$ can be further decomposed into three components and results in a quadruple decomposition of $\bA$ as described in \citet{liTheoreticalStudyDefinition2010a,liEvaluationVortexCriteria2014a,liEvaluationVortexCriteria2014b}.
The decomposition shows itself as a powerful framework for analyzing local motions of fluid elements and vortex criteria.

The NND proposed in \citet{liTheoreticalStudyDefinition2010a,liEvaluationVortexCriteria2014a,liEvaluationVortexCriteria2014b} is derived from a canonical form of real matrices discovered earlier by \citet{murnaghanCanonicalFormReal1931}, which is a counterpart over the field of real numbers to the famous Schur form for complex matrices \citep{schurUeberCharakteristischenWurzeln1909}.
In that paper, Isaai Schur generalized the previous results restricted to orthogonal \citep{stickelbergerUeberReelleOrthogonale1877} and hermitian \citep{autonneHermitien1901} matrices.
The canonical form of \citet{murnaghanCanonicalFormReal1931} is often referred to as the real Schur form.
It is known that both the complex and real Schur forms of a given matrix are not unique.
Actually, the NND is based on a special real Schur form among the possible variants.
However, \citet{liTheoreticalStudyDefinition2010a,liEvaluationVortexCriteria2014a,liEvaluationVortexCriteria2014b} didn't present a clarification of which variant of real Schur forms has been chosen and how to maintain uniqueness and consistency throughout the flow fields.

Recently, real and complex Schur forms as well as NND have received increasing attention.
For example,
\citet{keylockSyntheticVelocityGradient2017,keylockSchurDecompositionVelocity2018} employed the complex Schur form of velocity gradient tensor in their statistical model of fine structures in turbulence.
\citet{dasRevisitingTurbulenceSmallscale2020} analyzed fine structures of turbulence in the framework of NND.
\citet{hoffmanEnergyStabilityAnalysis2021} used  NND to analyze energy stability in incompressible turbulence.
\citet{zhu2021thermodynamic,zhu2021turbulence} derived a real Schur flow (RSF) as the compressible Taylor-Proudman limit in magnetohydrodynamics, for which the velocity gradient tensor takes a globally uniform real Schur form.
\citet{kronborg2022blood} applied NND to analyse the shear in blood flows.
\citet{Arun_Colonius_2024} analyze the collision of vortex rings with NND.
Just to mention a few.

Apart from the direct application of Schur forms and NND as mentioned above, there are also studies of fluid motions that are related to or similar in some sense to NND. For example, 
\citet{kolar2DVelocityFieldAnalysis2004,kolarVortexIdentificationNew2007} proposed a triple decomposition of motion (TDM) of the velocity gradient tensor, which is very similar to the quadruple NND proposed in \citet{liTheoreticalStudyDefinition2010a,liEvaluationVortexCriteria2014a,liEvaluationVortexCriteria2014b}.
Unfortunately, several works, such as \citet{dasRevisitingTurbulenceSmallscale2020,hoffmanEnergyStabilityAnalysis2021,kronborg2022blood,Arun_Colonius_2024}, misidentified the tools they used as TDM, when in fact they were NND.

\citet{kronborgTripleDecompositionVelocity2023} made an effort to clarify the relationship between TDM and the Schur forms.
But they overlooked the gap between the two; especially, they misidentified NND as TDM.
Also, they didn't reveal the intrinsic gap between the NNDs derived from real and complex Schur forms properly, and treated it merely as an issue of algorithmic convenience. 
Besides, they failed to characterize the non-uniqueness of these forms completely and to present a satisfactory standardization procedure, although they claimed so.

Considering the different types and non-uniqueness of Schur forms as well as the various possible definitions of NNDs, a clarification of their relationships is in demand, which is the aim of the current article.
We hope that this article will help eliminate the confusion that has existed in previous research and clear the way for further investigations and applications of Schur forms and NNDs in fluid mechanics and other areas.
The readers may also have an interest in the geometric or kinematical interpretations of Schur forms, which is out of the scope of this article.
We refer to \citet{zouSpiralStreamlinePattern2021}, which provides a nice discussion on streamline patterns of real Schur forms.

The structure of the article is as follows.
Since NND is based on a real Schur form, we first describe the Schur forms.
In \cref{sec:complex_Schur_form}, the general complex Schur form for a 3-by-3 real matrix is derived with a discussion of its non-uniqueness.
In \cref{sec:real_Schur_form}, the general real Schur form for a 3-by-3 real matrix is derived with a discussion of its non-uniqueness. A condition of uniqueness and the special real Schur form determined by this condition are proposed.
In \cref{sec:NND}, a brief discussion of normal and nilpotent matrices are presented firstly.
Based on the general complex Schur form, a construction of complex NND is presented.
Based on the special real Schur form proposed in \cref{sec:real_Schur_form}, several types of real NNDs are proposed.
Besides, we also proposed two normal-nonnormal decompositions.
Then, the distinction between the NNDs and the TDM proposed by \citet{kolar2DVelocityFieldAnalysis2004,kolarVortexIdentificationNew2007} is demonstrated.
In \cref{sec:gap}, we demonstrate that it is generally impossible to convert a complex NND into a real NND through unitary transformations, which implies that the two types of NNDs are nonequivalent.

\section{Complex Schur Form}
\label{sec:complex_Schur_form}

For later use, let's first restate a classical result about matrices.
Let $\bA\in\bbC^{n\times n}$ be a complex $n$-by-$n$ matrix.
A theorem of \citet{schurUeberCharakteristischenWurzeln1909} says that there is a unitary matrix $\bU$ such that $\tilbA=\bU\bA\bU^\dag$ is upper triangular, i.e., $\tilA_{ij}=0$ for all $i>j$, and the diagonal entries $\tilA_{ii}$'s are eigenvalues of $\bA$.
Denote the $i$-th row of $\bU$ as $\bmu_i$, then
\begin{align}
\label{eq:representation_formula}
    \bA = \bU^\dag\tilbA\bU = \sum_{i,j:i\le j} \tilA_{ij}\bmu_i^\dag\bmu_j.
\end{align}
The \cref{eq:representation_formula} is sometimes stated as a theorem and referred to as the \emph{representation theorem} of matrices.
\begin{theorem}
    Every matrix $\bA\in\bbC^{n\times n}$ has a representation as \cref{eq:representation_formula} in some orthonormal basis $\set{\bmu_i}$.
\end{theorem}
Indeed, it is an extension of the representation theorem of self-adjoint matrices, which says that every self-adjoint matrix has a diagonal representation in certain orthonormal basis.
An immediate consequence of the theorem is
\begin{corollary}
    In the representation formula \cref{eq:representation_formula}, $\bmu_n$ is a unit eigenvector of $\bA$.
\end{corollary}
Since $\set{\bmu_i}$ forms an orthonormal basis of $\bbC^n$, it is easy to see that $\bmu_n\bA=\tilA_{nn}\bmu_n$, i.e., $\bmu_n$ must be a unit eigenvector of $\bA$.
Without specification, we always refer to left eigenvectors.

Let $\bA\in\bbR^{3\times3}$ be a real 3-by-3 matrix.
It can be the component matrix of a second-order tensor in the basis aligned with the Cartesian coordinates $x,y,z$.
Since the characteristic polynomial of $\bA$ is cubic, $\bA$ always has a real eigenvalue $\lambda_3=\lambda_{\rr}$.
Let $\bme_{\rr}$ be a unit eigenvector associated with $\lambda_{\rr}$.
If $\bme_{\rr}$ is required to be real, then it has two possible choices, which are in opposite directions.
If it is allowed to be complex, then it can only be determined up to an arbitrary unitary factor $e^{\ii\phi_3}$.
Notice that we can always require that $\bme_{\rr}$ is a real vector. But we will not make this choice until necessary.

Fix $\bme_{\rr}$, let $\bmu,\bmv\in\bbC^3$ such that
\begin{align}
    \bU := 
    \begin{pmatrix}
        \bmu\\\bmv\\\bme_{\rr}
    \end{pmatrix}
\end{align}
is a unitary matrix.
There is
\begin{align}
\label{eq:general_real_Schur_form}
    \tilbA := \bU\bA\bU^\dag = 
    \begin{pmatrix}
    \tilbA_2 & \bmc^\dag \\
    \zero & \lambda_{\rr}
    \end{pmatrix}
\end{align}
with submatrices (i.e., blocks)
\begin{align}
    \tilbA_2 :=
    \begin{pmatrix}
        \bmu\bA\bmu^\dag & \bmu\bA\bmv^\dag \\
        \bmv\bA\bmu^\dag & \bmv\bA\bmv^\dag
    \end{pmatrix},\quad
    \bmc^\dag :=
    \begin{pmatrix}
        \bmu\bA\bme_{\rr}^\dag \\
        \bmv\bA\bme_{\rr}^\dag
    \end{pmatrix},
\end{align}
where $^\dag$ stands for adjoint (i.e., conjugate transpose).
The choices of $\bmu,\bmv$ are not unique.
If $\bV_2\in\bbC^{2\times2}$ is a unitary matrix, then $\bV\bU$ with
\begin{align}
    \bV:=
    \begin{pmatrix}
        \bV_2 & \zero \\
        \zero & 1
    \end{pmatrix}
\end{align}
will do the equivalent job as $\bU$ does, i.e., eliminate the first and second elements of the third row of $\bA$.
Conversely, any unitary transformation on $\bbC^3$ that leaves $\bme_{\rr}$ unchanged must have the form of $\bV$.

Besides of $\lambda_3=\lambda_{\rr}$, $\bA$ has another two eigenvalues $\lambda_1,\lambda_2\in\bbC$, which could be real or a pair of complex conjugate numbers.
They are also eigenvalues of $\tilbA_2$.
If $\lambda_1,\lambda_2\notin\bbR$, then $\bA$ can NOT be transformed into an upper triangular matrix by real orthogonal transformations.
With unitary transformations, however, $\bA$ can be converted to an upper triangular form.

Let
\begin{align}
    \bV_2 =
    \begin{pmatrix}
        \bma \\
        \bmb
    \end{pmatrix}
\end{align}
with $\bma,\bmb\in\bbC^2$
and
\begin{align}
    \tsup[2]{\bA}_2 := \bV_2\tilbA_2\bV_2^\dag =
    \begin{pmatrix}[1.5]
        \bma\tilbA_2\bma^\dag & \bma\tilbA_2\bmb^\dag \\
        \bmb\tilbA_2\bma^\dag & \bmb\tilbA_2\bmb^\dag
    \end{pmatrix}.
\end{align}
If $\tsup[2]{\bA}_2$ is upper triangular, then $\bmb$ must be a unit eigenvector of $\tilbA_2$.
Notice that $\tilbA_2$ is just the restriction of $\bA$ in the subspace orthogonal to $\bme_{\rr}$, which is independent of the choice of $\bmu,\bmv$.
Therefore, $\bmb$ only depends on $\bA$.
Again, $\bmb$ can only be determined up to an arbitrary unitary factor $e^{\ii \phi_2}$.
$\bma$ is a unit vector orthogonal to $\bmb$, i.e., $\bma\bma^\dag = 1$ and $\bma\bmb^\dag = 0$, hence $\bma$ is uniquely determined up to an arbitrary unitary factor $e^{\ii \phi_1}$.

Being an upper triangular matrix, $\bma\tilbA_2\bma^\dag$ and $\bmb\tilbA_2\bmb^\dag$ must be eigenvalues of $\tsup[2]{\bA}_2$ and $\tilbA_2$.
But their order is undetermined.
Let's denote $\bma\tilbA_2\bma^\dag = \lambda_1$ and $\bmb\tilbA_2\bmb^\dag = \lambda_2$.
There is
\begin{align}
    \tsup[2]{\bA} := \bV\tilbA\bV^\dag = 
    \begin{pmatrix}[1.5]
        \tsup[2]{\bA}_2 & \bV_2\bmc^\dag \\
        \zero & \lambda_{\rr}
    \end{pmatrix} =
    \begin{pmatrix}[1.5]
        \lambda_1 & \bma\tilbA_2\bmb^\dag & \bma\bmc^\dag \\
        0 & \lambda_2 & \bmb\bmc^\dag \\
        0 & 0 & \lambda_r
    \end{pmatrix},
\end{align}
which is a complex Schur form of $\bA$.
Notice that the complex Schur form is not unique.
Apart from permutations of diagonal entries, there are also undetermined unitary factors.
For all $\phi_1,\phi_2,\phi_3\in[0,2\pi)$, let
\begin{align}
    \bP := \diag\parentheses{e^{\ii \phi_1},e^{\ii \phi_2},e^{\ii \phi_3}}.
\end{align}
Then $\bP\bV\bU$ will do the equivalent job as $\bV\bU$ to transform $\bA$ into its complex Schur form.
Hence the \textit{general complex Schur form} of $\bA$ can be written as
\begin{align}
    \tsup[2]{\bA} = 
    \begin{pmatrix}[1.5]
        \lambda_1 & a_{12} e^{\ii (\phi_1-\phi_2)} & a_{13} e^{\ii (\phi_1-\phi_3)} \\
        0 & \lambda_2 & a_{23} e^{\ii (\phi_2-\phi_3)} \\
        0 & 0 & \lambda_{\rr}
    \end{pmatrix}.
\end{align}
Notice that only the differences between $\phi_1,\phi_2,\phi_3$ enter the results.
Let $\phi_{13}=\phi_1-\phi_3,\phi_{23}=\phi_2-\phi_3$, we obtain
\begin{align}
\label{eq:general_complex_Schur_form}
    \tsup[2]{\bA} = 
    \begin{pmatrix}[1.5]
        \lambda_1 & a_{12} e^{\ii (\phi_{13}-\phi_{23})} & a_{13} e^{\ii \phi_{13}} \\
        0 & \lambda_2 & a_{23} e^{\ii \phi_{23}} \\
        0 & 0 & \lambda_3
    \end{pmatrix}.
\end{align}
Since each of $\bma,\bmb,\bmc$ can be determined up to an arbitrary unitary factor, we can require that any two of $a_{12},a_{13},a_{23}$ are real.
Let's assume that $a_{13},a_{23}\in\bbR$.
But in general, $a_{12}$ is complex.
Assume that $a_{12}=\mu+\ii\nu$ with $\mu,\nu\in\bbR$.

If $\lambda_1,\lambda_2\in\bbR$, then the $\bU,\bV$ can be chosen as special orthogonal matrices, hence $a_{12}\in\bbR$.
Furthermore, we can set $\phi_{13}=m\pi,\phi_{23}=n\pi$, then
\begin{align}
\label{eq:general_complex_real_Schur_form}
    \tsup[2]{\bA} = 
    \begin{pmatrix}[1.5]
        \lambda_1 & a_{12} (-1)^{m-n} & a_{13} (-1)^m \\
        0 & \lambda_2 & a_{23} (-1)^n \\
        0 & 0 & \lambda_3
    \end{pmatrix}.
\end{align}
This is the \textit{real Schur form} of $\bA$ when $\lambda_1,\lambda_2\in\bbR$, which is a special case of \cref{eq:general_complex_Schur_form}.

\section{Real Schur Form}
\label{sec:real_Schur_form}
As mentioned at the end of \cref{sec:complex_Schur_form}, When $\bA\in\bbR^{3\times3}$ has three real eigenvalues, its general real Schur form is a particular case of its general complex Schur form, given in \cref{eq:general_complex_real_Schur_form}.
When $\bA$ has a pair of complex conjugate eigenvalues, it can not be transformed into an upper triangular matrix in the field of real numbers.
In this case, we can only have a block upper triangular matrix in the form of $\tilbA$ obtained in \cref{sec:complex_Schur_form}.
Especially, we can require that $\tilbA$ is real by restricting $\bU$ to orthogonal matrices.
It gives as the \emph{general real Schur form} of $\bA$ with a pair of complex conjugate eigenvalues.
But this leaves us one degree of freedom, which is the rotation in the subspace normal to $\bme_{\rr}$.
Therefore, an orthogonal transform can be applied to $\tilbA$ to meet one more requirement.
Let
\begin{align}
    \bQ = 
    \begin{pmatrix}
    \cos\varrho & \sin\varrho & 0 \\
    -\sin\varrho & \cos\varrho & 0 \\
    0 & 0 & 1
    \end{pmatrix},
\end{align}
then
\begin{align}
    \hatbA :=&\ \bQ\tilbA\bQ^\T \\
    =&\ \begin{pmatrix}
        \begin{split}
            \tilA_{11}\cos^2\varrho + \tilA_{22}\sin^2\varrho \\ 
            + (\tilA_{12}+\tilA_{21})\cos\varrho\sin\varrho  
        \end{split} 
        & 
        \begin{split}
            \tilA_{12}\cos^2\varrho - \tilA_{21}\sin^2\varrho \\
            + (\tilA_{22}-\tilA_{11})\cos\varrho\sin\varrho
        \end{split}
        & 
        \begin{split}
            c_1\cos\varrho + c_2\sin\varrho 
        \end{split}\\\\
        \begin{split}
            \tilA_{21}\cos^2\varrho - \tilA_{12}\sin^2\varrho \\
            + (\tilA_{22}-\tilA_{11})\cos\varrho\sin\varrho
        \end{split}
        &
        \begin{split}
            \tilA_{22}\cos^2\varrho + \tilA_{11}\sin^2\varrho \\
            - (\tilA_{12}+\tilA_{21})\cos\varrho\sin\varrho
        \end{split}
        & 
        \begin{split}
            c_2\cos\varrho - c_1\sin\varrho
        \end{split} \\\\
        \begin{split}
            0
        \end{split}
        &
        \begin{split}
            0
        \end{split}
        &
        \begin{split}
            \lambda_{\rr}
        \end{split}
    \end{pmatrix}.
\end{align}
We require that $\hatA_{11}=\hatA_{22}$, hence
\begin{align}
    (\tilA_{11}-\tilA_{22})\parentheses{\cos^2\varrho - \sin^2\varrho} + 2(\tilA_{12}+\tilA_{21})\cos\varrho\sin\varrho = 0,
\end{align}
which implies that
\begin{align}
    \tan2\varrho = \frac{\tilA_{22}-\tilA_{11}}{\tilA_{12}+\tilA_{21}}.
\end{align}
Since the period of $\tan$ is $\pi$, there are four values of $\varrho\in[0,2\pi)$ that satisfy the equation.
They determine two lines that are orthogonal to each other.
We can also obtain
\begin{align}
    \cos^2\varrho = \frac{1}{2}\parentheses{1\pm\frac{\tilA_{12}+\tilA_{21}}{|\tilbA_2|}},\quad
    \sin\varrho\cos\varrho = \pm\frac{\tilA_{22}-\tilA_{11}}{2|\tilbA_2|},
\end{align}
where 
\begin{align}
    |\tilbA_2|:=\sqrt{\norm{\tilbA_2}_F^2-2\det\tilbA_2}=\sqrt{\parentheses{\tilA_{12}+\tilA_{21}}^2 + \parentheses{\tilA_{22}-\tilA_{11}}^2}
\end{align}
is invariant under unitary transformations, hence the components of $\tilbA$ in its expression can be replaced by those of $\hatbA$.
With these values of $\varrho$, we have
\begin{align}
    &\hatA_{11} = \hatA_{22} = \frac{\tilA_{11}+\tilA_{22}}{2},\quad
    \hatA_{12} = \frac{\tilA_{12}-\tilA_{21}\pm|\tilbA_2|}{2},\\
    &\hatA_{21} = \frac{\tilA_{21}-\tilA_{12}\pm|\tilbA_2|}{2}.
\end{align}
Therefore,
\begin{align}
    \hatbA = 
    \begin{pmatrix}[2]
        \frac{\tilA_{11}+\tilA_{22}}{2} & \frac{\tilA_{12}-\tilA_{21}\pm|\tilbA_2|}{2} & \hatA_{13} \\
        \frac{\tilA_{21}-\tilA_{12}\pm|\tilbA_2|}{2} & \frac{\tilA_{11}+\tilA_{22}}{2} & \hatA_{23} \\
        0 & 0 & \lambda_{\rr}
    \end{pmatrix}.
\end{align}

Denote
\begin{align}
    \chi := \hatA_{11} = \hatA_{22},\quad
    \omega_3:=\hatA_{12}-\hatA_{21},\quad
    \gamma := \hatA_{12} + \hatA_{21} = \pm|\hatbA_2|.
\end{align}
The 2-by-2 diagonal block of $\hatbA$ can be written as
\begin{align}
    \hatbA_2 := 
    \begin{pmatrix}[1.5]
        \chi & \frac{\gamma + \omega_3}{2} \\
        \frac{\gamma - \omega_3}{2} & \chi
    \end{pmatrix},
\end{align}
We have mentioned that $\bme_{\rr}$ has two possible choices, which are in opposite directions.
If $\omega_3<0$, rotate the basis about the $\hatx$-axis for an angle of $\pi$.
The operation change the signs of $\omega_3$ and $\gamma$ simultaneously.
Therefore, the choice of $\bme_{\rr}$ is uniquely determined by requiring $\omega_3>0$.
We have also mentioned that $\varrho$ has four possible choices, which determine two lines that are orthogonal to each other.
If $\gamma<0$, rotate the basis about the $\hatz$-axis for an angle of $\pi/2$.
This operation changes the sign of $\gamma$ while keeping the sign of $\omega_3$.
Therefore, the choice of $\varrho$ is reduced to two opposite directions by requiring $\gamma>0$.
Denote
\begin{align}
    \hatA_{13} := -\beta, \quad
    \hatA_{23} := \alpha.
\end{align}
The two choices of $\varrho$ lead to different signs of $\alpha$ and $\beta$ simultaneously.
Therefore, the choice of $\varrho$ can be uniquely determined by assigning a sign for $\alpha$ or $\beta$.
We require that $\alpha>0$.
According to the discussions above, in the generic case that all of $\omega_3,\gamma,\alpha$ are nonzero,
\begin{align}
\label{eq:special_real_Schur_form}
    \hatbA = 
    \begin{pmatrix}[1.5]
        \chi & \frac{\gamma + \omega_3}{2} & -\beta \\
        \frac{\gamma - \omega_3}{2} & \chi & \alpha \\
        0 & 0 & \lambda_{\rr}
    \end{pmatrix}
\end{align}
can be uniquely determined by requiring
\begin{align}
    \omega_3>0,\quad \gamma>0,\quad \alpha>0.
\end{align}
Of course, this uniqueness condition is not the only choice.
For example, one can require $\omega_3<0$ instead.
But that only affects the appearance of the resulting standard matrix form and is merely a matter of convention.
If some of $\omega_3,\gamma,\alpha$ is zero, then the form of $\hatbA$ can be nonunique.
We should mention that \citet{kronborgTripleDecompositionVelocity2023} also noticed the nonuniqueness of Schur forms, but their consideration is restricted to a permutation of basis vectors, which is not essential to the forms and the construction of NNDs.

When $\bA$ has a pair of complex conjugate eigenvalues $\lambda_{1,2} = \lambda_{\rc\rr}\pm\ii\lambda_{\rc\ri}$, \cref{eq:special_real_Schur_form} is a special form of the real Schur forms of $\bA$.
In this case, there is
\begin{align}
    \lambda_{\rc\rr} = \chi,\quad
    \lambda_{\rc\ri} = \frac{1}{2}\sqrt{\omega_3^2-\gamma^2}.
\end{align}

We should emphasize that when $\bA$ has three real eigenvalues, the form of $\hatbA$ is still uniquely determined by the above procedure.
In other words, $\hatbA$ is a unique canonical form of $\bA$ even when $\bA$ has three real eigenvalues. 
But in this case, it is not a real Schur form of $\bA$.
In this case,
\begin{align}
    \lambda_{1,2} = \chi \pm \frac{1}{2}\sqrt{\gamma^2-\omega_3^2}.
\end{align}

\begin{remark}
In \cref{eq:special_real_Schur_form}, notice that
\begin{align}
    4\hatA_{12}\hatA_{21} = \gamma^2 - \omega_3^2
\end{align}
is the discriminant of the characteristic polynomial of $\hatbA_2$.
It can be used to determine whether $\bA$ has a pair of complex conjugate eigenvalues.
\end{remark}

\begin{remark}
    Since $\hatbA=(\bQ\bU)\bA(\bQ\bU)^\T$, $\bQ$ is a rotation matrix and $\bU$ can be chosen as a rotation matrix, therefore, the six nonzero components $\chi,\omega_3,\lambda_{\rr},\alpha,\beta,\gamma$ of $\hatbA$ are rotational invariants of $\bA$.
\end{remark}

\section{Normal-Nilpotent Decomposition}
\label{sec:NND}

\subsection{General Discussions}

A matrix $\bA\in\bbC^{n\times n}$ is said to be \emph{normal} if $\bA\bA^\dag=\bA^\dag\bA$.
$\bA$ is normal iff its complex Schur form is diagonal or its real Schur form is block diagonal.
$\bA$ is said to be \emph{nilpotent} if $\bA^n=0$.
$\bA$ is nilpotent iff it only has zero eigenvalues, iff its Schur form is strictly upper triangular.
Thus we can conclude that if $\bA$ is both normal and nilpotent, then $\bA=0$.
Therefore, it is very natural to decompose $\bA$ as a sum of a normal matrix and a nilpotent matrix by splitting the diagonal and off-diagonal parts of its complex Schur forms or splitting block diagonal and off-diagonal parts of its real Schur forms.
However, the splitting approach is not the only way to realize such a decomposition.
Actually, there may be more than one such decomposition for each Schur form of $\bA$.
Furthermore, considering the non-uniqueness of Schur forms, based on each variant of complex or real Schur forms of $\bA$, such a decomposition can be constructed.
All of them can be called \textit{normal-nilpotent decompositions} (NNDs).

There is another set of necessary and sufficient conditions for normal matrices, which may serve as guidance when constructing NNDs.
Let
\begin{align}
    \bS = \frac{\bA+\bA^\dag}{2},\quad
    \bW = \frac{\bA-\bA^\dag}{2}.
\end{align}
Then $\bA = \bS + \bW$ and $\bS^\dag=\bS, \bW^\dag=-\bW$.
It is evident that $\bA$ is normal iff $\bS\bW=\bW\bS$.
In the orthonormal basis composed of eigenvectors of $\bS$, suppose
\begin{align}
    \bS = 
    \begin{pmatrix}
        s_1 & 0 & 0 \\
        0 & s_2 & 0 \\
        0 & 0 & s_3
    \end{pmatrix},\quad
    \bW = 
    \begin{pmatrix}
        0 & w_3 & -w_2 \\
        -w_3 & 0 & w_1 \\
        w_2 & -w_1 & 0
    \end{pmatrix}.
\end{align}
Then
\begin{align}
    \bS\bW-\bW\bS =
    \begin{pmatrix}
        0 & w_3(s_1-s_2) & w_2(s_3-s_1) \\
        w_3(s_1-s_2) & 0 & w_1(s_2-s_3) \\
        w_2(s_3-s_1) & w_1(s_2-s_3) & 0
    \end{pmatrix},
\end{align}
which leads to the following
\begin{theorem}
\label{thm:normality_condition}
    The matrix $\bA\in\bbC^{3\times3}$ is normal only in one of the following three cases:
    \begin{enumerate}
        \item\label{item:distinct_s} All $s_1,s_2,s_3$ are distinct, $w_1=w_2=w_3=0$. In this case, $\bA$ is symmetric.
        \item\label{item:paired_s} $s_i=s_j\ne s_k$ for distinct $i,j,k$, $w_i=w_j=0$. In this case, $\bA$ is quasiorthogonal.
        \footnote{
        A matrix is said to be \emph{quasiorthogonal} if its columns are mutually orthogonal and so are its rows.
        But the columns and rows are not required to be normalized to unit length.
        The accurate name for this class of matrices should be \emph{orthogonal}, unfortunately, which has been widely accepted for matrices that should have been called \emph{orthonormal matrices}. }
        \item $s_1=s_2=s_3$. That is $\bS=(\tr\bA/3)\bI$.
    \end{enumerate}
\end{theorem}
Notice that the validity of these conditions does not depend on the diagonal form of $\bS$ since normality and nilpotency are independent of unitary transformations.
But the $s_i$'s and $w_i$'s refer to the matrix entries in the basis of eigenvectors of $\bS$.
This also justifies the procedure of constructing NND in the basis of Schur forms since they are generalizations of the basis of eigenvectors of self-adjoint matrices, as stated by the representation theorem \cref{eq:representation_formula}.

\subsection{Complex NND}
Let $\tsup[2]{\bLambda}$ be the diagonal part of $\tsup[2]{\bA}$ given in \cref{eq:general_complex_Schur_form} and $\tsup[2]{\bDelta}$ be the off-diagonal part.
That is
\begin{align}
\label{eq:complex_NND}
    \tsup[2]{\bLambda} = 
    \begin{pmatrix}[1.5]
        \lambda_1 & 0 & 0 \\
        0 & \lambda_2 & 0 \\
        0 & 0 & \lambda_3
    \end{pmatrix},\quad
    \tsup[2]{\bDelta} = 
    \begin{pmatrix}[1.5]
        0 & a_{12} e^{\ii (\phi_{13}-\phi_{23})} & a_{13} e^{\ii \phi_{13}} \\
        0 & 0 & a_{23} e^{\ii \phi_{23}} \\
        0 & 0 & 0
    \end{pmatrix}.
\end{align}
Then we have $\tsup[2]{\bA} = \tsup[2]{\bLambda} + \tsup[2]{\bDelta}$.
Obviously, $\tsup[2]{\bLambda}$ is normal and $\tsup[2]{\bDelta}$ is nilpotent.
In the generic case, all $\lambda_1,\lambda_2,\lambda_3$ are distinct from each other.
In the same sprite of the \cref{item:distinct_s} of \cref{thm:normality_condition}, the normal part should contains no off-diagonal entries.
In particular, we can not add an anti-symmetric matrix to $\tsup[2]{\bLambda}$.
Besides, we can not add a symmetric matrix to $\tsup[2]{\bLambda}$ either, otherwise it will violate the nilpotency of $\tsup[2]{\bDelta}$.
Hence the above construction of NND is the only natural one in this case.
Of course, the decomposition is complex, and $\tsup[2]{\bDelta}$ is not unique.
When $\bA$ has three real eigenvalues, $\tsup[2]{\bLambda}$ is real and we can make $\tsup[2]{\bDelta}$ real too, as shown by \cref{eq:general_complex_real_Schur_form}.
When $\bA$ has a pair of complex conjugate eigenvalues, $\tsup[2]{\bLambda}$ can be converted to a real matrix through a unitary transformation, but $\tsup[2]{\bDelta}$ can not be converted to a real matrix by the same unitary transformation.
The details are presented in \cref{sec:gap}.

The NND used by \citet{keylockSyntheticVelocityGradient2017,keylockSchurDecompositionVelocity2018} is of this type.
It has several advantages, such as
\begin{enumerate}
    \item The Schur form has a simple upper triangular form with eigenvalues on its diagonal.
    \item There is only one natural way of decomposing the Schur form into the sum of a normal part and a nilpotent part.
    \item The normal part and the nilpotent part have no overlap, resulting in a clean decomposition.
    \item The cases of real and complex conjugate eigenvalues are treated in a unified way.
\end{enumerate}
As a mathematical model of physical quantities such as the velocity gradient tensor, the complex NND is adequate.
However, the physical interpretation of the complex matrices is not clear yet.
In contrast, real NNDs can have a much clearer physical meaning, as demonstrated in the literature.

\subsection{Real NND}

In \citet{liTheoreticalStudyDefinition2010a}, the real NND of $\bA$ with three eigenvalues is a special case of the complex NND as constructed above.
When $\bA$ has a pair of complex conjugate eigenvalues, the real NND is constructed based on the special real Schur form \cref{eq:special_real_Schur_form}.
However, in both the real and complex eigenvalue cases, those are not the only ways to construct real NNDs.

Let's consider the case of complex eigenvalues first.
Since the special real Schur form \cref{eq:special_real_Schur_form} has two identical diagonal entries, the normal part of $\hatbA$ corresponds to the \cref{item:paired_s} of \cref{thm:normality_condition}.
In particular, the normal part of $\hatbA$ can only have off-diagonal entries in the positions of $12$ and $21$.
We have three natural ways to decompose the off-diagonal part of the submatrix $\hatbA_2$,
which are:
\begin{enumerate}
    \item Quasiorthogonal-nilpotent,
    \item Symmetric-nilpotent,
    \item Symmetric-antisymmetric.
\end{enumerate}
Each type of decomposition above leads to a type of decomposition of $\hatbA$.
Since both symmetric and antisymmetric matrices are normal, the first two ways of decomposition belong to NND and the third does NOT.

Denote
\begin{align}
    \phi := \frac{\omega_3 + \gamma}{2},\quad
    \psi := \frac{\omega_3 - \gamma}{2}.
\end{align}
The decomposition
\begin{align}
    \hatbA = \hatbN + \hatbGamma
\end{align}
can be realized in the following ways:
\begin{outline}[enumerate]
    \1 \textbf{Quasiorthogonal-nilpotent decomposition.}
    The normal component $\hatbN$ contains the diagonal of $\hatbA$ and the anti-symmetric part of $\hatbA_2$, hence is orthogonal.
    There are two natural ways to separate an anti-symmetric part from $\hatbA_2$, which leads to two variants of NND: The one minimizes $\|\hatbN\|_F$, the other maximizes $\|\hatbN\|_F$. 
    The $\|\hatbGamma\|_F$ in the two approaches are the same.
    In both cases, $\hatbN$ is normal and $\hatbGamma$ is nilpotent.
        \2 \textbf{Minimize the normal part:}
        \begin{align}
        \label{eq:quasiorthogonal-nilpotent_decomposition_minN}
            \hatbN :=
            \begin{pmatrix}
                \chi & \psi & 0 \\
                -\psi & \chi & 0 \\
                0 & 0 & \lambda_{\rr}
            \end{pmatrix},\quad
            \hatbGamma :=
            \begin{pmatrix}
                0 & \gamma & -\beta \\
                0 & 0 & \alpha \\
                0 &  0 & 0
            \end{pmatrix}.
        \end{align}
        In this case, $\|\hatbN\|_F$ is minimized because $|\gamma-\omega_3|<\gamma+\omega_3$.
        This variant of NND is the one proposed in \citet{liTheoreticalStudyDefinition2010a}.
        
        \2 \textbf{Maximize the normal part:} 
        \begin{align}
        \label{eq:quasiorthogonal-nilpotent_decomposition_maxN}
            \hatbN :=
            \begin{pmatrix}
                \chi & \phi & 0 \\
                -\phi & \chi & 0 \\
                0 & 0 & \lambda_{\rr}
            \end{pmatrix},\quad
            \hatbGamma :=
            \begin{pmatrix}
                0 & 0 & -\beta \\
                \gamma & 0 & \alpha \\
                0 &  0 & 0
            \end{pmatrix}.
        \end{align}
        In this case, $\|\hatbN\|_F$ is maximized because the off-diagonal entry of $\hatbA_2$ with a larger absolute value is kept in $\hatbN$.

    \1 \textbf{Symmetric-nilpotent decomposition.}
    The normal component $\hatbN$ contains the diagonal of $\hatbA$ and the symmetric part of $\hatbA_2$.
    Similarly, there are two natural ways to separate a symmetric part from $\hatbA_2$, which leads to two variants of NND:
        \2 \textbf{Minimize the normal part:}
        \begin{align}
        \label{eq:symmetric-nilpotent_decomposition_minN}
            \hatbN :=
            \begin{pmatrix}
                \chi & -\psi & 0 \\
                -\psi & \chi & 0 \\
                0 & 0 & \lambda_{\rr}
            \end{pmatrix},\quad
            \hatbGamma :=
            \begin{pmatrix}
                0 & \omega_3 & -\beta \\
                0 & 0 & \alpha \\
                0 &  0 & 0
            \end{pmatrix}.
        \end{align}
        
        \2 \textbf{Maximize the normal part:} 
        \begin{align}
        \label{eq:symmetric-nilpotent_decomposition_maxN}
            \hatbN :=
            \begin{pmatrix}
                \chi & \phi & 0 \\
                \phi & \chi & 0 \\
                0 & 0 & \lambda_{\rr}
            \end{pmatrix},\quad
            \hatbGamma :=
            \begin{pmatrix}
                0 & 0 & -\beta \\
                -\omega_3 & 0 & \alpha \\
                0 &  0 & 0
            \end{pmatrix}.
        \end{align}

    \1 \textbf{Normal-nonnormal decompositions.}
    The $\hatbN$ contains the diagonal of $\hatbA$ and the symmetric part or antisymmetric part of $\hatbA_2$. 
    In both cases, $\hatbN$ is normal but $\hatbGamma$ is generally not nilpotent.
    Hence the resulting decompositions of $\hatbA$ are NOT NNDs.
        \2 \textbf{Symmetric-nonnormal:}
        \begin{align}
            \hatbN :=
            \begin{pmatrix}[1.5]
                \chi & \frac{\gamma}{2} & 0 \\
                \frac{\gamma}{2} & \chi & 0 \\
                0 & 0 & \lambda_{\rr}
            \end{pmatrix},\quad
            \hatbGamma :=
            \begin{pmatrix}[1.5]
                0 & \frac{\omega_3}{2} & -\beta \\
                -\frac{\omega_3}{2} & 0 & \alpha \\
                0 &  0 & 0
            \end{pmatrix}.
        \end{align}
        \2 \textbf{Quasiorthogonal-nonnormal:}
        \begin{align}
            \hatbN :=
            \begin{pmatrix}[1.5]
                \chi & \frac{\omega_3}{2} & 0 \\
                -\frac{\omega_3}{2} & \chi & 0 \\
                0 & 0 & \lambda_{\rr}
            \end{pmatrix},\quad
            \hatbGamma :=
            \begin{pmatrix}[1.5]
                0 & \frac{\gamma}{2} & -\beta \\
                \frac{\gamma}{2} & 0 & \alpha \\
                0 &  0 & 0
            \end{pmatrix}.
        \end{align}
\end{outline}

\begin{remark}
    When $\bA$ has three real eigenvalues, \cref{eq:special_real_Schur_form} is not a real Schur form of $\bA$, but it is still a uniquely determined canonical form of $\bA$.
    In this case,
    \begin{align}
        \chi = \frac{\lambda_1 + \lambda_2}{2}
    \end{align}
    and $|\gamma|\ge|\omega_3|$, hence $\psi\le0$.
    The decompositions constructed above based on the canonical form \cref{eq:special_real_Schur_form} extends to this case.
\end{remark}

\begin{remark}
    \citet{kolar2DVelocityFieldAnalysis2004,kolarVortexIdentificationNew2007} defined a triple decomposition of motion (TDM) earlier than NND of \citet{liTheoreticalStudyDefinition2010a,liEvaluationVortexCriteria2014a,liEvaluationVortexCriteria2014b}, but the two decompositions have very similar appearances, which has caused some confusion.
    Here we point out several distinctions between the two:
    \begin{enumerate}
        \item The \emph{purely asymmetric tensor} in \citet{kolar2DVelocityFieldAnalysis2004,kolarVortexIdentificationNew2007}, defined by the vanishing of the product of each pair of off-diagonal entries, is generally NOT nilpotent.
        For example, the matrix
        \begin{align}
            \begin{pmatrix}
                0 & \gamma & 0 \\
                0 & 0 & \alpha \\
                \beta & 0 & 0
            \end{pmatrix}
        \end{align}
        is purely asymmetric according to \citet{kolar2DVelocityFieldAnalysis2004,kolarVortexIdentificationNew2007}, but its determinant is nonzero, indicating nonzero eigenvalues. Hence it is not nilpotent.
        Of course, the purely asymmetric tensor can be nilpotent in several special cases, especially in the 2D case. 
        But the concept behind it is different from the nilpotent tensor.
        Actually, a condition of nilpotency can be phrased as \emph{a given $n$-by-$n$ matrix is nilpotent iff among any $n$ entries that occupy $n$ rows and $n$ columns, there is a zero.}

        \item TDM maximizes the norm of the purely asymmetric tensor through changing of basis, while the NNDs do not maximize the nilpotent tensor, neither do they choose basis according to optimization principles.

        \item Regardless of the difference in their choice of basis, TDM and NND produce different results.
        For example, suppose the \emph{basic reference frame} (BRF) of \citet{kolar2DVelocityFieldAnalysis2004,kolarVortexIdentificationNew2007} is the same as the basis of the special real Schur form \cref{eq:special_real_Schur_form}.
        When $\bA$ has complex conjugate eigenvalues, its TDM coincides with \cref{eq:quasiorthogonal-nilpotent_decomposition_minN};
        but when $\bA$ has three real eigenvalues, its TDM coincides with \cref{eq:symmetric-nilpotent_decomposition_minN} instead.
        Of course, we can define a new type of NND by combining \cref{eq:quasiorthogonal-nilpotent_decomposition_minN} and \cref{eq:symmetric-nilpotent_decomposition_minN} in each case to include the results of TDM, but it is somewhat artificial and we don't know whether this is always the case.
    \end{enumerate}
\end{remark}
\section{Gap between Complex and Real NNDs}
\label{sec:gap}

In this section, we show that it is generally impossible to transform a complex NND into a real NND.
Consider the general complex Schur form \cref{eq:general_complex_Schur_form}.
Its complex NND is just the splitting of the diagonal and off-diagonal parts, as given by \cref{eq:complex_NND}.
Let's find a unitary transformation that converts it to a real NND.
Since $\lambda_3=\lambda_{\rr}$ is real, the required unitary transformation should leave $\bme_{\rr}$ unchanged, hence is essentially a unitary transformation in the subspace normal to $\bme_{\rr}$.

Denote
\begin{align}
    \bLambda_2 := \diag\parentheses{\lambda_1,\lambda_2},
\end{align}
where $\lambda_{1,2}=\lambda_{\rc\rr}\pm\ii\lambda_{\rc\ri}$ is a pair of complex conjugate numbers.
We are seeking a unitary transformation that converts $\bLambda_2$ into a real matrix.
Let
\begin{align}
    \bR_2 := \frac{1}{\sqrt{2}}
    \begin{pmatrix}
        -\ii & 1 \\
        1 & -\ii
    \end{pmatrix}.
\end{align}
It can be verified that $\bR_2$ is unitary and 
\begin{align}
    \bN_2 := \bR_2\bLambda_2\bR_2^\dag = 
    \begin{pmatrix}
        \lambda_{\rc\rr} & \lambda_{\rc\ri} \\
        -\lambda_{\rc\ri} & \lambda_{\rc\rr}
    \end{pmatrix},
\end{align}
which is real.
Of course, to transform $\bLambda_2$ into a real matrix, the choice of $\bR_2$ is not unique.
For every real orthogonal matrix
\begin{align}
    \bQ_2 =
    \begin{pmatrix}
        \cos\theta & \sin\theta \\
        -\sin\theta & \cos\theta
    \end{pmatrix},
\end{align}
$\bQ_2\bR_2$ will also do the job.
Let 
\begin{align}
    \bR := 
    \begin{pmatrix}
        \bR_2 & \zero \\
        \zero & 1
    \end{pmatrix},\quad
    \bQ :=
    \begin{pmatrix}
        \bQ_2 & \zero \\
        \zero & 1
    \end{pmatrix}
\end{align}
and $\tsup[2]{\bA}$ be the general complex Schur form \cref{eq:general_complex_Schur_form}, then
\begin{align}
    \tsup[3]{\bA} := (\bQ\bR)\tsup[2]{\bA}(\bQ\bR)^\dag =
    \begin{pmatrix}
        \bM & \bmd^\dag \\
        \zero & \lambda_{\rr}
    \end{pmatrix},
\end{align}
where
\begin{align}
    \bM := 
    \begin{pmatrix}[2]
        \lambda_{\rc\rr} - \frac{\ii}{2}a_{12}e^{\ii(2\theta+\phi_{13}-\phi_{23})} & \lambda_{\rc\ri} + \frac{1}{2}a_{12}e^{\ii(2\theta+\phi_{13}-\phi_{23})} \\
        -\lambda_{\rc\ri} + \frac{1}{2}a_{12}e^{\ii(2\theta+\phi_{13}-\phi_{23})} & \lambda_{\rc\rr} + \frac{\ii}{2}a_{12}e^{\ii(2\theta+\phi_{13}-\phi_{23})}
    \end{pmatrix}
\end{align}
and
\begin{align}
    \bmd := \parentheses{
        \frac{\ii}{\sqrt{2}}a_{13}e^{-\ii(\theta+\phi_{13})} + \frac{1}{\sqrt{2}}a_{23}e^{\ii(\theta-\phi_{23})},
        \frac{1}{\sqrt{2}}a_{13}e^{-\ii(\theta+\phi_{13})} + \frac{\ii}{\sqrt{2}}a_{23}e^{\ii(\theta-\phi_{23})}}.
\end{align}
That's all we can do.
Now we need to determine $\phi_{13},\phi_{23}$ and $\theta$ such that $\tsup[3]{\bA}$ is real.
We rewrite $\bM = \bN_2 + \xi\bL_2$, where $\xi:=\frac{1}{2}a_{12}e^{\ii(2\theta+\phi_{13}-\phi_{23})}$ and
\begin{align}
    \bL_2 :=
    \begin{pmatrix}
        -\ii & 1 \\
        1 & \ii
    \end{pmatrix}.
\end{align}
However, whatever $\phi_{13},\phi_{23}$ and $\theta$ are chosen, $\bM$ can never be real, so can $\tsup[3]{\bA}$.
Therefore, a complex NND can NOT be converted into a real NND by unitary transformations in general.
They are intrinsically distinct decompositions.

\section{Conclusions and Discussions}

The major contributions of this article are:
\begin{enumerate}
    \item The general complex and real Schur forms are derived constructively, with special consideration for their non-uniqueness. 
    Conditions of uniqueness are found to obtain standardized Schur forms.
    
    \item Conditions for normality and nilpotency are presented. Based on the general complex Schur form, a complex NND is constructed.
    Based on a special real Schur form, several real NND as well as normal-nonnormal decompositions are constructed.
    A comparison of advantages and disadvantages between complex and real NNDs is presented.

    \item Several confusing points are explained, including the distinction between NND and the TDM proposed by \citet{kolar2DVelocityFieldAnalysis2004,kolarVortexIdentificationNew2007}, the intrinsic gap between complex and real NNDs.
\end{enumerate}
We hope that the objective of the article, i.e., clarifying the confusion about Schur forms and NNDs existing in research, has been achieved.

Now the open question is, provided so many versions of decompositions, which is the right one to use?
Typically, there are two widely adopted usages:
\begin{enumerate}
    \item Use complex Schur form and complex NND throughout.
    \item Use the special block real Schur form \cref{eq:special_real_Schur_form} and correspongding NND when $\bA$ has complex conjugate eigenvalues; use the real Schur form \cref{eq:general_complex_real_Schur_form} and the corresponding NND when $\bA$ has three real eigenvalues.
\end{enumerate}
But now one has an additional option:
Use the special block real Schur form \cref{eq:special_real_Schur_form} and the corresponding NND throughout, i.e., for both real and complex eigenvalue cases.
Of course, the above description didn't take into account the different variants of real NNDs as proposed in \cref{sec:NND}, which provides even more options. 
These options will serve different needs.

\printbibliography

@article{Arun_Colonius_2024, 
title={Velocity gradient analysis of a head-on vortex ring collision}, 
volume={982},
journal={Journal of Fluid Mechanics}, 
author={Arun, Rahul and Colonius, Tim}, 
year={2024}, 
pages={A16}
}

@inproceedings{autonneHermitien1901,
  title = {Sur l’hermitien},
  booktitle = {Rendiconti del Circolo Matematico di Palermo (1884-1940)},
  author = {Autonne, M. Léon},
  date = {1901-08-25},
  volume = {16},
  pages = {104--128},
  langid = {italian},
  language = {Italian}
}

@article{dasRevisitingTurbulenceSmallscale2020,
  title = {Revisiting Turbulence Small-Scale Behavior Using Velocity Gradient Triple Decomposition},
  author = {Das, Rishita and Girimaji, Sharath S.},
  date = {2020-06},
  journaltitle = {New Journal of Physics},
  shortjournal = {New J. Phys.},
  volume = {22},
  number = {6},
  pages = {063015},
  publisher = {IOP Publishing},
  langid = {english}
}

@article{hoffmanEnergyStabilityAnalysis2021,
  title = {Energy Stability Analysis of Turbulent Incompressible Flow Based on the Triple Decomposition of the Velocity Gradient Tensor},
  author = {Hoffman, Johan},
  date = {2021-08-01},
  journaltitle = {Physics of Fluids},
  volume = {33},
  number = {8},
  pages = {081707},
  langid = {english}
}

@article{keylockSyntheticVelocityGradient2017,
  title = {Synthetic Velocity Gradient Tensors and the Identification of Statistically Significant Aspects of the Structure of Turbulence},
  author = {Keylock, Christopher J.},
  date = {2017-08-23},
  journaltitle = {Physical Review Fluids},
  shortjournal = {Phys. Rev. Fluids},
  volume = {2},
  number = {8},
  pages = {084607},
  publisher = {American Physical Society}
}

@article{keylockSchurDecompositionVelocity2018,
  title = {The Schur Decomposition of the Velocity Gradient Tensor for Turbulent Flows},
  author = {Keylock, Christopher J.},
  date = {2018-08},
  journaltitle = {Journal of Fluid Mechanics},
  volume = {848},
  pages = {876--905},
  publisher = {Cambridge University Press},
  langid = {english}
}

@inproceedings{kolar2DVelocityFieldAnalysis2004,
  title = {2D Velocity-Field Analysis Using Triple Decomposition of Motion},
  booktitle = {Proceedings of the Fifteenth Australasian Fluid Mechanics Conference},
  author = {Kolář, Václav},
  date = {2004-12},
  publisher = {The University of Sydney},
  eventtitle = {The Fifteenth Australasian Fluid Mechanics Conference}
}

@article{kolarVortexIdentificationNew2007,
  title = {Vortex Identification: New Requirements and Limitations},
  shorttitle = {Vortex Identification},
  author = {Kolář, Václav},
  date = {2007-08-01},
  journaltitle = {International Journal of Heat and Fluid Flow},
  shortjournal = {International Journal of Heat and Fluid Flow},
  series = {Including Special Issue of Conference on Modelling Fluid Flow (CMFF’06), Budapest},
  volume = {28},
  number = {4},
  pages = {638--652},
  langid = {english}
}

@ARTICLE{kronborg2022blood,
AUTHOR={Kronborg, Joel  and Svelander, Frida  and Eriksson-Lidbrink, Samuel  and Lindström, Ludvig  and Homs-Pons, Carme  and Lucor, Didier  and Hoffman, Johan },
TITLE={Computational Analysis of Flow Structures in Turbulent Ventricular Blood Flow Associated With Mitral Valve Intervention},
JOURNAL={Frontiers in Physiology},
VOLUME={13},
YEAR={2022},
}

@article{kronborgTripleDecompositionVelocity2023,
  title = {The Triple Decomposition of the Velocity Gradient Tensor as a Standardized Real Schur Form},
  author = {Kronborg, Joel and Hoffman, Johan},
  date = {2023-03-01},
  journaltitle = {Physics of Fluids},
  volume = {35},
  pages = {031703}
}

@thesis{liTheoreticalStudyDefinition2010a,
  title = {Theoretical Study on the Definition of Vortex},
  author = {Li, Zhen},
  date = {2010},
  institution = {Tsinghua University},
  location = {Beijing, China},
  langid = {chinese},
  language = {Chinese}
}

@article{liEvaluationVortexCriteria2014a,
  title = {Evaluation of vortex criteria by virtue of the quadruple decomposition of velocity gradient tensor},
  author = {Li, Zhen and Zhang, Xiwen and He, Feng},
  date = {2014-03-05},
  journaltitle = {Acta Physica Sinica},
  shortjournal = {Acta Phys. Sin.},
  volume = {63},
  number = {5},
  pages = {054704--054704},
  langid = {chinese},
  language = {Chinese}
}

@online{liEvaluationVortexCriteria2014b,
  title = {Evaluation of Vortex Criteria by Virtue of the Quadruple Decomposition of Velocity Gradient Tensor},
  author = {Li, Zhen and Zhang, Xiwen and He, Feng},
  date = {2014},
  eprint = {2406.02558},
  eprinttype = {arXiv},
  eprintclass = {physics},
  pubstate = {prepublished}
}

@article{murnaghanCanonicalFormReal1931,
  title = {A Canonical Form for Real Matrices under Orthogonal Transformations},
  author = {Murnaghan, F. D. and Wintner, A.},
  date = {1931-07-01},
  journaltitle = {Proceedings of the National Academy of Sciences},
  shortjournal = {PNAS},
  volume = {17},
  number = {7},
  pages = {417--420},
  publisher = {National Academy of Sciences},
  langid = {english}
}

@article{schurUeberCharakteristischenWurzeln1909,
  title = {Über die charakteristischen Wurzeln einer linearen Substitution mit einer Anwendung auf die Theorie der Integralgleichungen},
  author = {Schur, Isaai},
  date = {1909-12-01},
  journaltitle = {Mathematische Annalen},
  shortjournal = {Math. Ann.},
  volume = {66},
  number = {4},
  pages = {488--510},
  langid = {ngerman},
  language = {German}
}

@inproceedings{stickelbergerUeberReelleOrthogonale1877,
  title = {Ueber reelle orthogonale Substitutionen},
  author = {Stickelberger, Ludwig},
  date = {1877},
  publisher = {Schweizerische Nationalbibliothek},
  location = {Bern},
  langid = {German}
}

@article{zhu2021turbulence,
    author = {Zhu, Jian-Zhou},
    title = "{Compressible helical turbulence: Fastened-structure geometry and statistics}",
    journal = {Physics of Plasmas},
    volume = {28},
    number = {3},
    pages = {032302},
    year = {2021},
    month = {03},
}

@article{zhu2021thermodynamic,
    author = {Zhu, Jian-Zhou},
    title = "{Thermodynamic and vortic structures of real Schur flows}",
    journal = {Journal of Mathematical Physics},
    volume = {62},
    number = {8},
    pages = {083101},
    year = {2021},
    month = {08},
}

@article{zouSpiralStreamlinePattern2021,
  title = {Spiral Streamline Pattern around a Critical Point: Its Dual Directivity and Effective Characterization by Right Eigen Representation},
  shorttitle = {Spiral Streamline Pattern around a Critical Point},
  author = {Zou, Wennan and Xu, Xiangyang and Tang, Changxin},
  date = {2021-06-01},
  journaltitle = {Physics of Fluids},
  volume = {33},
  number = {6},
  pages = {067102},
  publisher = {American Institute of Physics}
}
\end{document}